\documentclass[journal]{IEEEtran}%
\IEEEoverridecommandlockouts
\usepackage{soul}
\usepackage{cite}
\usepackage{diagbox}
\usepackage{amsmath,amssymb,amsfonts}
\usepackage{amsthm,bm}
\usepackage{mathrsfs}
\usepackage{colortbl}
\usepackage{hhline}
\usepackage{tabularx}
\usepackage{enumerate}
\usepackage{booktabs}
\usepackage{booktabs}% http://ctan.org/pkg/booktabs
\newcommand{\tabitem}{~~\llap{\textbullet}~~}
\usepackage{multicol}
\usepackage{graphicx}
\usepackage{subfigure} 
\usepackage{textcomp}
\usepackage{url}
\usepackage{stfloats}
\usepackage{mathtools}
\usepackage{cuted}
\usepackage{array}  %表格上下左右居中
\usepackage{makecell}
\usepackage {threeparttable}
\usepackage{tablefootnote}
\usepackage{multirow}
\usepackage{colortbl}
\def\BibTeX{{\rm B\kern-.05em{\sc i\kern-.025em b}\kern-.08em
    T\kern-.1667em\lower.7ex\hbox{E}\kern-.125emX}}

\usepackage{stackengine}
\usepackage{enumerate}
\usepackage{booktabs}
\usepackage{subfig}
\usepackage{multicol}
\usepackage[noend]{algpseudocode}
\usepackage[table,xcdraw]{xcolor}
\usepackage{pgfplots}
\usepackage{pgfplotstable}
\usepackage[ruled,linesnumbered]{algorithm2e} 
\SetKwRepeat{Do}{do}{while}
\definecolor{Reds}{RGB}{0,0,0}%{0,47,200}%{26,42,255}%{5,39,175}40,69,122
\definecolor{blue}{RGB}{0,0,0}
\usepackage{setspace}
\usepackage{color}
\usepackage{tabularray}
\definecolor{Iron}{rgb}{0.811,0.815,0.815}
\makeatletter
\def\BState{\State\hskip-\ALG@thistlm}
\makeatother

\algnewcommand\algorithmicforeach{\textbf{for each}}
\algdef{S}[FOR]{ForEach}[1]{\algorithmicforeach\ #1\ \algorithmicdo}
\begin{document}

\title{Streamlined Transmission: A Semantic-Aware XR Deployment Framework Enhanced by Generative AI

}
\author{Wanting Yang, Zehui Xiong, \textit{Senior Member, IEEE}, Tony Q. S. Quek, \textit{Fellow, IEEE}, Xuemin Shen, \textit{Fellow, IEEE}\thanks{
Wanting Yang, Zehui Xiong and Tony Q. S. Quek are with the Pillar
of Information Systems Technology and Design, Singapore University of
Technology and Design, Singapore (e-mail: wanting\_yang@sutd.edu.sg; zehui\_xiong@sutd.edu.sg; tonyquek@sutd.edu.sg);

Xuemin Shen is with the Department of Electrical and Computer Engineering, University of Waterloo, Waterloo, ON N2L 3G1, Canada (e-mail:
sshen@uwaterloo.ca).

}
}

\makeatletter
\setlength{\@fptop}{0pt}
\makeatother

\maketitle

\vspace{-1.8cm}
\begin{abstract}
In the era of 6G, featuring compelling visions of digital twins and metaverses, Extended Reality (XR) has emerged as a vital conduit connecting the digital and physical realms, garnering widespread interest. Ensuring a fully immersive wireless XR experience stands as a paramount technical necessity, demanding the liberation of XR from the confines of wired connections. In this paper, we first introduce the technologies applied in the wireless XR domain, delve into their benefits and limitations, and highlight the ongoing challenges. We then propose a novel deployment framework for a broad XR pipeline, termed “GeSa-XRF”, inspired by the core philosophy of Semantic Communication (SemCom) which shifts the concern from “how” to transmit to “what” to transmit. Particularly, the framework comprises three stages: data collection, data analysis, and data delivery. In each stage, we integrate semantic awareness to achieve streamlined transmission and employ Generative Artificial Intelligence (GAI) to achieve collaborative refinements. For the data collection of multi-modal data with differentiated data volumes and heterogeneous latency requirements, we propose a novel SemCom paradigm based on multi-modal fusion and separation and a GAI-based robust superposition scheme. To perform a comprehensive data analysis, we employ multi-task learning to perform the prediction of field of view and personalized attention and discuss the possible preprocessing approaches assisted by GAI. Lastly, for the data delivery stage, we present a semantic-aware multicast-based delivery strategy aimed at reducing pixel level redundant transmissions and introduce the GAI collaborative refinement approach. The performance gain of the proposed GeSa-XRF is preliminarily demonstrated through a case study.
\end{abstract}


\newtheorem{definition}{Definition}
\newtheorem{lemma}{Proposition}
\newtheorem{theorem}{Theorem}

\newtheorem{property}{Property}

\vspace{0cm}
\section{Introduction}

With the compelling visions of metaverses and digital twins, extended reality (XR), as a bridge between the real and virtual worlds, has garnered significant attention.
To ensure a fully immersive experience, eliminating the tether of XR stands as a paramount technical imperative. {\color{blue}Given the inherent scarcity of radio resources,} there exists an urgent necessity for a novel communication framework customized for XR. {\color{blue}Opportunely, the emergence of semantic communication offers fresh insights to address this requirement.}  

{\color{blue}The} pivotal concept in SemCom is shifting from the traditional concern of ``how" to transmit information to the consideration of ``what" to transmit, {\color{blue}thus streamlining 
the transmission}~\cite{yang2022semantic}.  While numerous review articles highlight the substantial benefits of SemCom to XR scenarios, {\color{blue}the comprehensive implementation methods for the entire XR pipeline, encompassing data collection, analysis, and delivery, have been notably deficient~\cite{yang2022semantic,wang2023adaptive}}. {\color{blue}The existing monolithic SemCom frameworks struggle to flexibly meet the varying data volumes, data modalities, and heterogeneous quality of service requirements.}
{\color{blue} Furthermore, for the most mature deep learning (DL)-based SemCom, the unavoidable error floor during the training inevitably 
results in the degradation of the user's immersive experience in reconstructed virtual scene at the mobile XR devices~\cite{qin2021semantic}.} 

Fortunately, the advent of generative artificial intelligence (GAI) has ushered in a promising opportunity to revolutionize the SemCom framework{\color{blue}\cite{grassucci2023generative, raha2023generative}}. In the paradigm shift from SemCom to generative SemCom, the extraction of semantic information evolves into the determination of the PROMPT, providing enhanced flexibility in communication system design, while alleviating transmission burdens~\cite{yang2023semantic}. {\color{blue}Furthermore, the diffusion model, renowned as one of the foremost GAI technologies due to its exceptional ability in generating high-quality images, shows considerable potential for enhancing high-definition XR experiences. Nonetheless, this advancement comes with the drawback of prolonged inference latency, particularly when tasked with generating content of substantial data volumes.} {\color{blue}Given that} XR itself is a computationally sensitive application,  the wholesale transfer of the existing {\color{blue}generative} SemCom into XR  undoubtedly brings more challenges to instant virtual scene rendering. As such, the strategic integration of GAI into the XR pipeline, capitalizing on its strengths to enhance overall XR performance while minimizing {\color{blue}computational} burdens, becomes a thoughtful imperative.

{\color{blue}To fill the research gaps}, we  systematically
{\color{blue}review} the currently available technologies, delve into
their benefits and limitations, and highlight the ongoing
challenges in the XR domain. Based on this,
we propose a Semantic-aware XR Deployment Framework Enhanced by GAI, termed ``GeSa-XRF".
{\color{blue}In contrast to the existing purely poin  t-to-point SemCom framework, GeSa-XRF is structured around three distinct functional tasks, each addressing critical aspects across the stages of data collection, analysis, and delivery. Leveraging the innovative potential of SemCom and GAI, semantic awareness and GAI techniques are seamlessly integrated into each task. The specific contributions are outlined as follows:}
\begin{itemize}
    % \item  For the data collection stage, {\color{blue}different from the available multi-modal SemCom aimed at decision-making~\cite{Qinzhijin2023}, we propose a novel SemCom paradigm based on multi-modal fusion and separation for the reconstruction of multi-modal signals. Moreover, instead of choosing between SemCom and traditional communication~\cite{wang2023adaptive}, we exploit the powerful robustness of SemCom
    % and propose a GAI-based superposition scheme to deal with the collection of multi-modal data with differentiated data volumes and heterogeneous latency requirements.}
\item {\color{blue}In the data collection stage, distinct from the existing multi-modal SemCom approach targeted at decision-making~\cite{Qinzhijin2023}, we introduce a novel SemCom paradigm centered on multi-modal fusion and separation for reconstructing multi-modal signals. Furthermore, rather than opting between SemCom and traditional communication~\cite{wang2023adaptive}, we harness the robustness of SemCom and propose a GAI-based superposition scheme to manage the collection of multi-modal data with varied data volumes and heterogeneous latency requirements,  with a  trade-off among  performance,
inference latency, and training complexity.}
    \item {\color{blue}In the data analysis stage, considering the limitations of computing resources on mobile devices, we employ the tile-based rendering method. Subsequently, by leveraging multi-task learning (MTL) techniques, we develop a unified algorithm for attention assessment and field of view (FoV) prediction. Utilizing the attention assessment results, we assign varying degrees of semantic significance to different tiles. Moreover, we classify the tiles into foreground and background segments based on their semantic importance. For the background tiles, we employ GAI-based proactive preprocessing guided by the FoV prediction results to enhance the overall XR performance.}
    \item {\color{blue}In the data delivery stage, we incorporate the semantic significance obtained during the data analysis stage into the delivery strategy to reduce pixel-level redundant transmissions. Specifically, the strategy involves semantic-aware multicast cluster decision, semantic-aware transcoding, and semantic-aware scheduling for the foreground tiles. Simultaneously, we introduce a three-dimensional multi-user quality of experience (QoE) evaluation metrics, encompassing weighted resolution based on tile significance, playback smoothness, and playback synchronization to guide strategy optimization. Furthermore, our case study delves into semantic-aware cluster decision-making and transcoding, primarily showcasing the performance improvements facilitated by GeSa-XRF.}
\end{itemize}

\begin{figure*}
    \centering \includegraphics[width=1\linewidth]{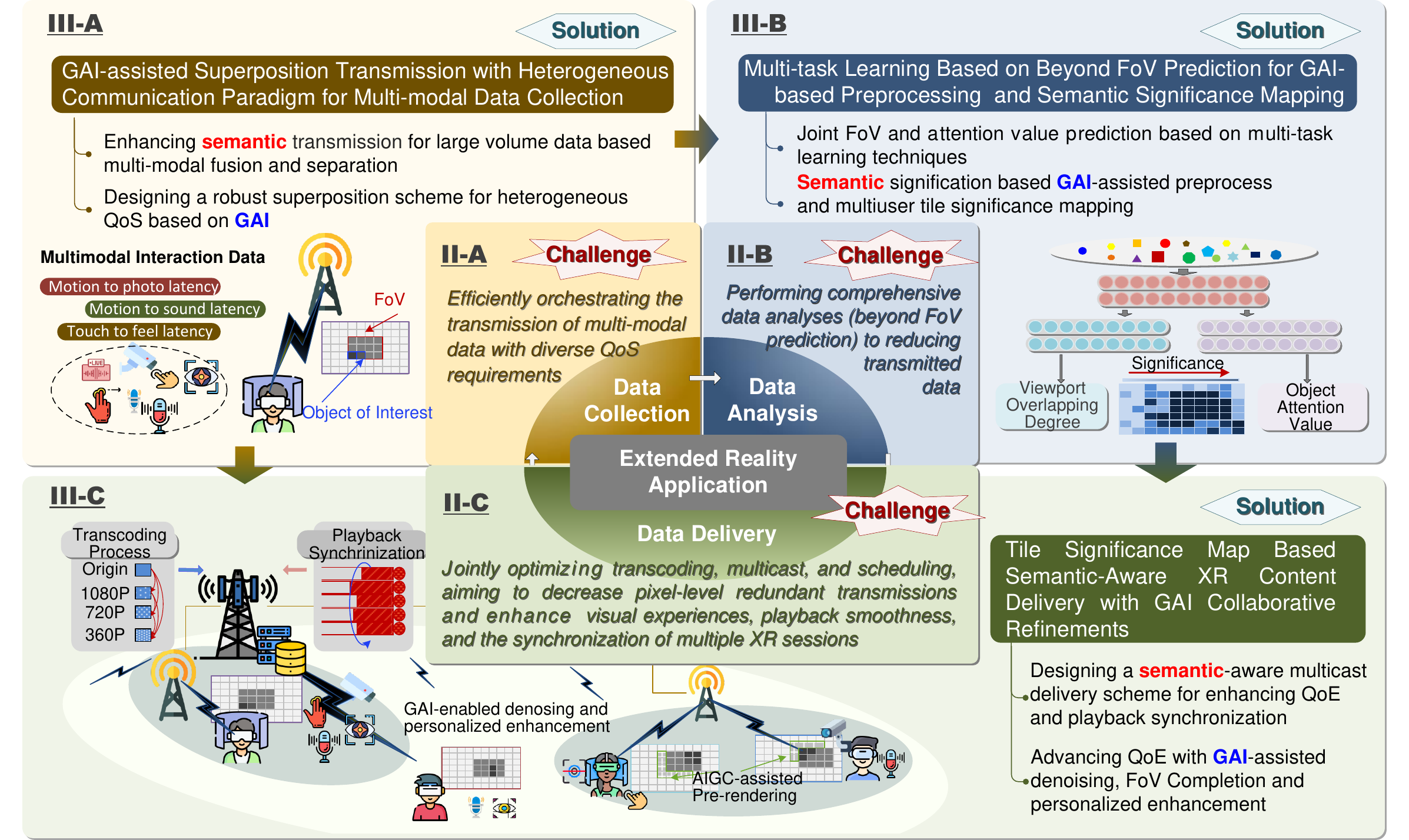}
    \caption{Overview of GeSa-XRF Deployment.}
    \label{fig:Overview}
\end{figure*}

\section{XR Deployment Overview and Challenges}
To maintain generality, we focus on a broad three-stage XR deployment pipeline. The available techniques and the challenges in each stage are discussed below:

  \begin{table*}
 \small
 \centering
 \caption{{Summary of the Role of Semantic Awareness and GAI in GeSa-XRF.}}
 \begin{threeparttable}
 \begin{tabular}{|p{0.04cm} p{0.1cm}|p{1.2cm}|p{4.65cm}|p{4.65cm}|p{4.65cm}|}
  \hline
  \multicolumn{3}{|c|}{\diagbox{Stage}{Tech.}} & \multicolumn{1}{c|}{\bfseries Semantic Awareness} & \multicolumn{1}{c|}{ $\bm \leftarrow$ \bfseries Relationship  $\bm \to$ }& \multicolumn{1}{c|}{ \bfseries Generative AI} \\
  \hline \hline
\multirow{4}{*}{\raisebox{-5.2\normalbaselineskip}[0pt][0pt]{\rotatebox{90}{{{Stage I {*}}}}} } & \multirow{4}{*}{\raisebox{-6.7\normalbaselineskip}[0pt][0pt]{\rotatebox{90}{{{Data Collection}}}} } & {\quad}  \newline \vspace{ -0.6cm} \newline {\color{blue}Objective} & \multicolumn{3}{l|} { \raisebox{-0.2\normalbaselineskip}[0pt][0pt]{ \makecell[l]{  \color{blue}\textit{Efficiently orchestrating the transmission of multi-modal data with diverse QoS requirements}}}} \\ \cline{3-6}
& &\raisebox{-2\normalbaselineskip}[0pt][0pt]{\color{blue}Benefit} & {{\quad}} \newline \vspace{ -0.6cm} \newline \tabitem  Achieving efficient multi-modal semantic compression \newline \tabitem Enhancing the robustness of superposition scheme & {\quad}  \newline \vspace{ -0.6cm} \newline  Utilizing GAI to further enhance the active adversarial capability of the robust semantic decoder in the superposition framework \newline \vspace{-0.2cm}& {\quad} \newline \vspace{ -0.6cm}\newline \tabitem Empowering the decoder to proactively capture and eliminate interference caused by superposition on ongoing transmissions \\ \cline{3-4} \cline{6-6}
& & {\quad}  \newline \vspace{ -0.6cm} \newline Approach & {\quad}  \newline \vspace{ -0.6cm} \newline DL-based SemCom paradigm  \newline \vspace{-0.5cm} & {  $\qquad 
\quad \xleftarrow[{\qquad \qquad \qquad \quad}]{}$ }& {\quad}  \newline \vspace{ -0.6cm} \newline {\color{blue}GAI-basd denoising}
\\\cline{3-6}
& & {\quad}  \newline \vspace{ -0.65cm} \newline {\color{blue}Challenge} & \multicolumn{3}{l|} { \makecell[l]{ \quad \vspace{-0.2cm}\\ 
\tabitem \color{blue}{The requirement to explore the explainability of semantic containers, determine the optimal location of output  } \\ {\color{blue} semantic features, and improve the performance of multimodal fusion and separation is crucial.}\\
\tabitem {\color{blue}
The necessity for pre-training an effective discriminator to effectively evaluate the quality of generated data is}\\ {\color{blue}  vital for enhancing the training stability during fine-tuning of GANs.}}} \\ 

 \hline \hline
\multirow{4}{*}{\raisebox{-5.7\normalbaselineskip}[0pt][0pt]{\rotatebox{90}{{{Stage II}}}} } & \multirow{4}{*}{\raisebox{-7.3\normalbaselineskip}[0pt][0pt]{\rotatebox{90}{{{ Data Analysis}}}} } 
& {\quad}  \newline \vspace{ -0.6cm} \newline {\color{blue}Objective} & \multicolumn{3}{l|} { \raisebox{-0.2\normalbaselineskip}[0pt][0pt]{\color{blue}\textit{Performing comprehensive data analyses (beyond FoV prediction) to reducing 
transmitted 
data}}} \\ \cline{3-6}
& &\raisebox{-2.5\normalbaselineskip}[0pt][0pt]{\color{blue}Benefit} & {{\quad}} \newline \vspace{ -0.6cm} \newline \tabitem  By linking semantic features within the tiles of the FoV to the user's attention for each type of semantic feature, varying importance can be assigned to individual tiles  & {\quad}  \newline \vspace{ -0.6cm} \newline  By analyzing the semantic features of  FoV and  user's preferences, the tile can be personalized into background and foreground ones for processing separately \newline \vspace{-0.2cm}& {\quad} \newline \vspace{ -0.6cm}\newline \tabitem For the background tiles, the GAI-based predictive rendering techniques can reduce the number of tiles to be transmitted in both the requested FoV as and the next FoV \\ \cline{3-4} \cline{6-6}
& & {\quad}  \newline \vspace{ -0.6cm} \newline Approach & {\quad}  \newline \vspace{ -0.6cm} \newline {\color{blue}Joint training based MTL}   \newline \vspace{-0.5cm} & {  $\qquad 
\quad \xrightarrow[{\qquad \qquad \qquad \quad}]{}$ }& {\quad}  \newline \vspace{ -0.6cm} \newline Image inpainting and outpainting  \\ \cline{3-6}
& & {\quad}  \newline \vspace{ -0.65cm} \newline {\color{blue}Challenge} & \multicolumn{3}{l|} { \makecell[l]{ \quad \vspace{-0.2cm}\\ 
\tabitem \color{blue}{Designing the MTL neural network structure delicately to maximize the sharing of underlying features and      } \\ {\color{blue} conserve computational resources, all while ensuring prediction accuracy, presents a significant challenge. }\\  
\tabitem {\color{blue}The varied computing capabilities necessitate personalized trade-offs between computing latency for   }\\ {\color{blue}inpainting and outpainting, as well as transmission latency.}}} \\ 

\hline \hline
\multirow{3}{*}{\raisebox{-8\normalbaselineskip}[0pt][0pt]{\rotatebox{90}{{{Stage III}}}} } & \multirow{3}{*}{\raisebox{-9.5\normalbaselineskip}[0pt][0pt]{\rotatebox{90}{{{ Data Delivery}}}} } 
& {\quad}  \newline \vspace{ -0.7cm} \newline {\color{blue}Objective} & \multicolumn{3}{l|} {\makecell[l] {\quad \vspace{-0.3cm} \\\textit {\color{blue}Jointly optimizing transcoding, multicast, and scheduling, aiming to decrease pixel-level redundant transmissions} \\\textit {\color{blue}and enhance  visual experiences,  playback smoothness, and the synchronization of multiple XR sessions}}} \\ \cline{3-6}
& &\raisebox{-2.8\normalbaselineskip}[0pt][0pt]{\color{blue}Benefit} & {{\quad}} \newline \vspace{ -0.6cm} \newline \tabitem  Achieve pixel-level redundancy removal in resource-constrained wireless transmission, by integrating semantic awareness into transcoding and multicast   & {\quad}  \newline \vspace{ -0.6cm} \newline Leveraging GAI techniques to collaboratively enhance semantic-aware content delivery, incorporating denoising, FoV completion, and personalized enhancements\newline \vspace{-0.2cm}& {\quad} \newline \vspace{ -0.6cm}\newline \tabitem Denoising received distorted tiles\newline \tabitem Recovering the dropped tiles exceeding latency requirements \newline \tabitem Enabling personalized enhancements and refinement\\ \cline{3-4} \cline{6-6}
& & {\quad}  \newline \vspace{ -0.22cm} \newline Approach & {\quad}  \newline \vspace{ -0.6cm} \newline \tabitem Combinatorial optimization \newline \tabitem  Decision decomposition based on  GAI \& large language model    & {\quad} \newline \vspace{0.4cm}{  $\qquad 
\quad \xleftarrow[{\qquad \qquad \qquad \quad}]{}$ }& {\quad}  \newline \vspace{ -0.6cm} \newline \tabitem Image recovery \newline \tabitem Image inpainting \newline \tabitem real-time translation  \\
\cline{3-6}
& & {\quad}  \newline \vspace{ -0.7cm} \newline {\color{blue}Challenge} & \multicolumn{3}{l|} { \makecell[l]{ \quad \vspace{-0.2cm}\\ 
\tabitem \color{blue}{  Decoupling the high-dimensional multi-constraint optimization problem on joint semantic-aware  multicast } \\ {\color{blue} cluster decision, transcoding, and scheduling poses a significant challenge. }\\  
\tabitem {\color{blue}The optimization and offloading of computing consumption and latency for GAI-based collaborative   }\\ {\color{blue} refinements is paramount. }}} \\
\hline

 \end{tabular}

 \begin{tablenotes}
\item[*] {\scriptsize It should be noted that at each stage we focus only on the most significant issues. The superposition scheme for uplink data collection proposed in Stage I can also be applied to the multicast-based downlink transmission process in Stage III. }
\end{tablenotes}

  \end{threeparttable}
   \label{Comparison}
\end{table*}

\begin{figure*}[t]
    \centering \includegraphics[width=0.95\linewidth]{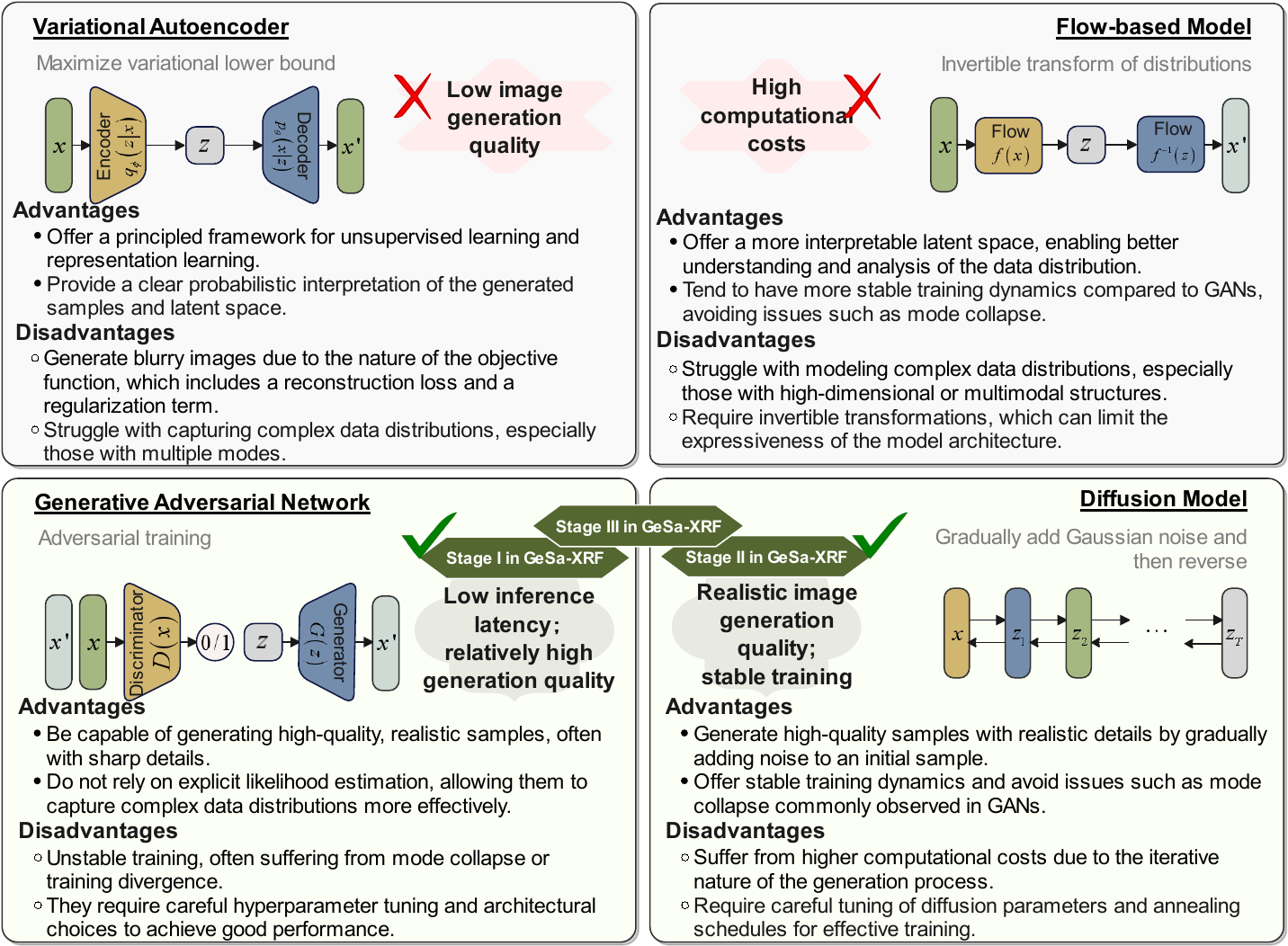}
    \caption{\color{blue} Summary of mainstream GAI technologies~\cite{zhang2023generative} }
    \label{fig:gai}
\end{figure*}
\subsection{Triggering Stage: Data Collection}
\label{sec: II3}
The effectiveness of immersive interactions hinges upon the integration of multi-sensory data  exhibiting in distinct modalities with varying data volumes. For instance, data streams like  audio and video for environment mapping may require data rates exceeding 100 Mbps. Conversely, data volumes for eye tracking and haptic commands are often just a few bytes. Furthermore, human brain reaction times to visual, auditory, and tactile stimuli differ significantly~\footnote{Public research suggests that motion-to-photon latency should be under 15–20 ms for a seamless experience. Motion-to-sound latency can be tolerated up to 30 ms, while touch-to-feel latency imposes the most stringent requirements, often as low as 1 ms. }. %Due to the large data volume of downlink feedback data, the permissible latency of the uplink is curtailed to varying degrees. 

Given the sporadic and random nature of the user behavior, allocating a dedicated frequency bandwidth or semi-persistent scheduling for these data may result in low bandwidth utilization and the weak capability of latency guarantee. To this end, the puncturing scheme and superposition scheme~\cite{darabi2022hybrid} posited for scenarios with the coexistence of enhanced mobile broadband (eMBB) and  ultra-reliable and low-latency communications (URLLC) traffic appears to be more suitable for the targeted XR case. Nevertheless, the adoption of such a joint scheduling mechanism inevitably impacts the transmission performance of eMBB, leading to the degradation of the quality of virtual scene construction.  Thus, \textit{efficiently orchestrating the transmission of multi-modal data with diverse latency requirements still stands  as the foremost challenge in enhancing the XR user experience and realizing its success.}

% Within the  network slicing architecture, to reduce latency in requesting resources for URLLC traffic, dedicating frequency bandwidth is a common solution, but it leads to low bandwidth utilization in XR due to sporadic data generation, and also impose challenges for other sensory data collection. Meanwhile, semi-persistent scheduling falls short of meeting latency requirements due to the stochastic nature of user behavior. 
 
\subsection{Intermediate Stage: Data Analysis}
\label{sec:IIB}
Despite understanding user interactions, data analysis can also play an essential role in enhancing the XR performance.  Nowadays, the FoV prediction has evolved into an integral technology for wireless XR transmission. However, most of the ongoing refinements  only contain eye-tracking data. While some advanced research has integrated video saliency detection into FoV predictions~\cite{li2023spherical}, it is important to clarify that view direction is influenced not only by the distribution of attractive objects but also by personal preferences. Nonetheless, few studies establish a connection between individual interests and video content, potentially limiting the generalizability of existing models.

Furthermore, while accurate FoV predictions can effectively reduce the latency experienced by users by proactively transmission, this approach fails to address the issues of limited radio resources in fact, especially in multi-user scenarios. As such, relying solely on FoV predictions proves insufficient. Drawing inspiration from the fundamental concept of SemCom,  we believe that optimizing XR sessions should involve strategies aimed at reducing transmitted data volume.  In this sense, \textit{there arises a need to explore reliable and comprehensive data analysis (beyond FoV prediction), with the overarching objective of further reducing transmitted data, all while maintaining the user's immersive experience.}

\subsection{Feedback Stage: Data Delivery }
To enhance the immersive XR environment, timely feedback of multi-sensory data to users is crucial.  Among the multiple types of sensory data, visual data is particularly data-intensive, which is always perceived as a formidable obstacle for wireless mobile VR deployments. In recent literature, the utilization of online transcoding emerges as a prominent technique for enhancing user experience~\cite{zhong2022multi}. However,  the user's attention degree to different tiles in the requested FoV is always neglected and all the tiles are treated uniformly. 
This oversight results in existing efforts optimizing only up to the point of ensuring video playback smoothness and failing to optimize user experience in a personalized manner. 

Meanwhile, for multi-user scenarios, despite the distinct regions of interest for each user, there is considerable overlap of the FoVs requested in a session~\cite{zhong2022multi}. Given the inherent broadcast nature of wireless networks, multicast presents itself as a promising efficient transmission technology. While this method is straightforward to implement in a single cell, determining clusters for multicast in the typical heterogeneous network (HetNet) becomes challenging, which is influenced by both the available bandwidth of individual base stations (BSs) and the user distribution. Additionally, due to the lack of consideration for the significance of each tile, current multicast strategies solely address the transmission redundancy at the tile level. Hence, \textit{there is a need to explore a strategy aiming to decrease pixel-level redundant transmissions while accommodating personalized visual experiences, playback smoothness, and the synchronization of multiple XR sessions.}

\section{ Exploring GeSa-XRF: Illustrative Insights}
In this section, the specific implementation details of GeSa-XRF in each stage are presented, respectively, where the main roles of semantic awareness and GAI are summarized in Table~\ref{Comparison}. {\color{blue}Meanwhile, the meanstream GAI techniques have been presented in Fig.~\ref{fig:gai}~\footnote{The diagrammatic representation of the model in this figure is from \url{https://lilianweng.github.io/posts/2021-07-11-diffusion-models/}}.}

\subsection{Multi-modal Data Collection}
In this subsection, we focus on the orchestration of the multi-modal data collection with diverse latency requirements, which is  visualized in Fig.~\ref{fig:task1}.
\subsubsection{Designing a Multi-modal SemCom Framework for Large Volume Data }
\label{sec:IIIA1}
First, we target sensory data with large volumes, such as audio and video.  {\color{blue}Considering the ultra-high latency requirements for uplink transmission}, the DL-based multi-modal SemCom stands out the first-choice~\cite{Qinzhijin2023}, which allows not only for more efficient semantic compression by removing inter-modal redundancies, but also for cross-modal parsing  at the XR server to enhance decoding robustness {\color{blue}with low inference latency}.

\begin{figure*}[t]
    \centering \includegraphics[width=0.95\linewidth]{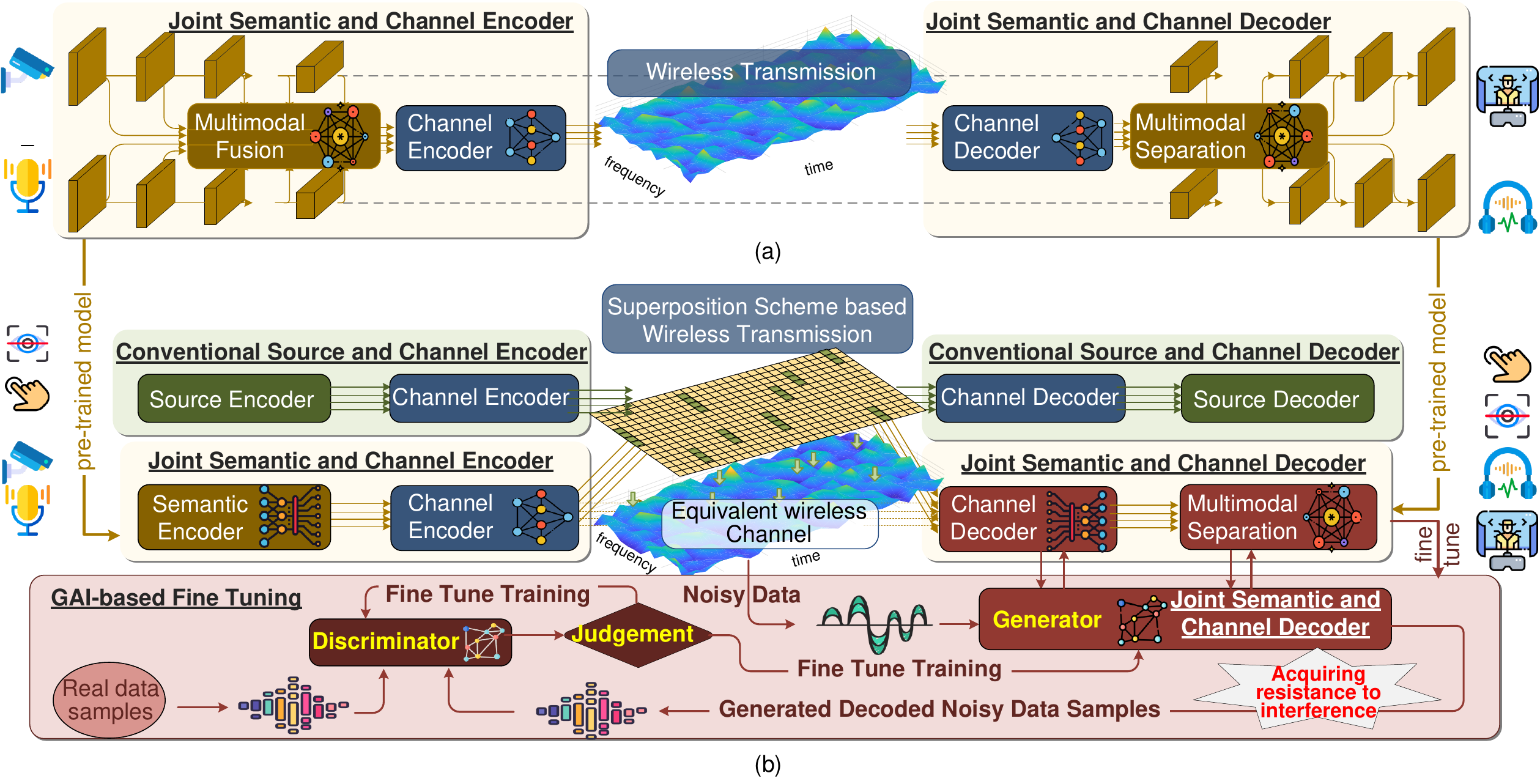}
    \caption{\color{blue}GAI-assisted superposition transmission with heterogeneous
communication paradigm for multi-modal data collection. (a) DL-based multi-modal SemCom framework; (b) Robust GAI-assisted superposition scheme for heterogeneous transmission.  }
    \label{fig:task1}
\end{figure*}
Different from the available multi-modal SemCom aimed at decision-making, e.g., visual question answering, the XR server targets at the reconstruction of multi-modal signals. 
To this end, a multi-modal SemCom framework based on multi-modal fusion and separation is proposed as shown in Fig.~\ref{fig:task1}(a). Inspired by the DL-based SemCom framework designed for uni-modal data reconstruction, the semantic encoders and decoders herein adopt a symmetric structure, both of which consist of two integral components. One corresponds to the neural network for reconstructing individual uni-modal data, where the encoder part acts as a layer-structured semantic container, while the decoder part functions as a cascading semantic reasoner. The other relates to the multi-modal fusion or separation module. At the transmitter, the fusion module processes the semantic features output by certain layers of the individual semantic containers, facilitating semantic fusion to eliminate redundant information. Conversely, the separation module at the receiver takes the fused data, parsing multiple semantic features, and conveys them to the corresponding semantic reasoners for incremental enhancement of semantic inference layer by layer. 
Moreover, the realization of an effective fusion and separation pair necessitates the identification of appropriate semantic features, implying an optimal location for output semantic features within the semantic container.

\subsubsection{Designing a GAI-assisted Superposition Scheme for Coexistence of Multi-modal Data}
According to existing literature~\cite{qin2021semantic}, {\color{blue}DL-based} SemCom exhibits greater robustness compared to conventional bit-based communications and can achieve high-quality data reconstruction even at low signal-to-noise ratios. Therefore, the  {\color{blue}multi-model SemCom proposed above} holds the promise to endow the superposition with the resilience to withstand inter-user interference. With this in mind, we conceive  a novel superposition scheme for the coexistence of multi-modal sensory data as shown in Fig.~\ref{fig:task1}(b). 

Specifically, the data with small data volume, like haptic information, employ  conventional communications to mitigate the computational latency, which are transmitted to the XR server by overlaying the semantic transmission described in section~\ref{sec:IIIA1}.  Such the superposition scheme can be identified as power-domain non-orthogonal multiple access. Due to the end-to-end training fashion, it is important to note that the decoding of the semantic signal cannot involve a direct extraction from the superimposed semantic and bit signals~\cite{mu2023exploiting}.  In this sense, at the destination, this scheme follows the ``bits-to-semantics successive interference cancellation ordering", as bit communications require no prior training.
Furthermore,  to strengthen the ability of SemCom to cope with superposition interference, {\color{blue}we can employ GAI to fine-tune the semantic decoder, with relatively low training complexity. Considering its single forward pass characteristics and relative high generation quality, we resort to}  the generative adversarial network (GAN) to further refine the joint channel and semantic decoder, avoiding introducing much extra latency. Therein, the pre-trained joint channel and semantic decoder {\color{blue}in Section~\ref{sec:IIIA1}} is treated as the initial generator in GAN. Simultaneously, to augment the decoder's active denoising capabilities, a discriminator is introduced.
 % which encounters two types of inputs. One is the distorted output of the generator, and the other is the high-quality data samples, which are labeled lower and higher scores, respectively. 
 The learning object of the discriminator is to distinguish the distorted data and the high-quality data. Against this, the learning objective of the generator is to deceive the discriminator to get a higher score. In other words, the well-trained discriminator can act as a fitting process of implicit functions used to guide semantic decoding optimization, which can reflect the relationship between the compressed semantic information and the quality of the reconstructed data. 

\subsection{Multi-task Data Analysis}
In this subsection, we delve deeper into the potential of the data analysis stage to streamline the transmission and improve the overall performance of the XR.
\subsubsection{Joint FoV and Attention Value Prediction based on Multi-task Learning Techniques}
As discussed in Section~\ref{sec:IIB}, the personalized interests of the users are ignored in existing FoV predictions, which causes the FoV prediction model to be retrained for every new XR video.
 To address this concern, we propose to introduce the factor of  user-object-attention (UOA) value into the FoV prediction, which can lay the foundation for  personalized video saliency prediction without the need for retraining.
Meanwhile, according to our prior work~\cite{du2023attention},  the prediction of UOA values also holds a crucial role in assigning semantic significance to tiles within the FoV, which assists in optimizing transcoding, multicasting, and personalized experience.
However, our initial approach in UOA prediction solely relies on the user-object-attention level dataset for Metaverse research. In fact, the shift of the FoV can also serve as an indicator of the user's attention to various objects.  In other words, FoV prediction and user-object-attention prediction are intricately connected tasks.

 Consequently, we resort to the utilization of MTL techniques~\cite{samant2022framework} to train a model capable of simultaneously performing both prediction tasks, leveraging shared information to enhance overall prediction performance as shown in Fig.~\ref{fig:task2}(a). To establish a suitable dataset, a mapping that delineates the correspondence between objects and tiles within the FoV needs to be created. This enables the fusion of the two datasets dedicated to each prediction task, facilitating multi-task training. Additionally, to recognize potential shared underlying patterns between the two tasks, the neural network architecture should incorporate shared layers to extract low-level common features, which is followed by two task-specific blocks for each prediction task capturing unique patterns related to individual tasks. To maintain an overall optimization objective, we propose the creation of a joint loss function that combines losses from both tasks. The expansion on how the prediction results enhance XR session performance will be explored in Section~\ref{sec:IIIB2}.
 \subsubsection{Semantic signification based GAI-assisted preprocess and multi-user tile significance mapping}
 \label{sec:IIIB2}
\begin{figure*}
    \centering \includegraphics[width=0.95\linewidth]{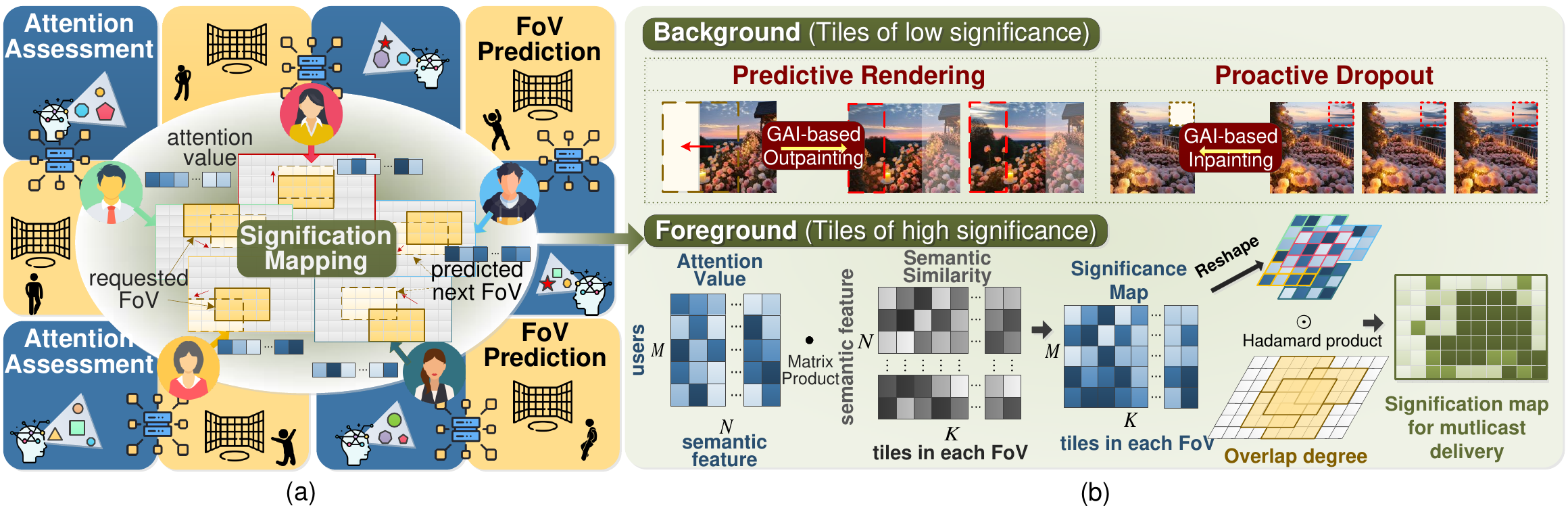}
    \caption{Multi-task learning based on beyond FoV prediction for GAI-based preprocessing and semantic significance mapping. (a) Joint FoV prediction and attention assessment based on MTL; (b) Preprocessing for background and foreground tiles. }
    \label{fig:task2}
\end{figure*}

Based on the inclusion or exclusion of objects of interest to the user, we propose to categorize tiles into two types for separate processes as shown in Fig.~\ref{fig:task2}(b).                                                
\begin{itemize}
    \item \textbf{Background tile}: {\color{blue}Background tiles used to create the backdrop or distant scenery.}    
    Due to substantial overlap between adjacent FoVs, devices can employ advanced GAI techniques to conduct outpainting for predictive rendering of background information {\color{blue}in the unit of tile in advance. Considering the more relaxed downlink latency requirements resulting from proactive pre-processing, we favour the choice of the diffusion model, e.g., the stable diffusion given its excellent image generation details.}
    Herein, the XR server only needs to transmit the FoV prediction results to the users, which can enhance the smoothness of user experiences during FoV transitions without consuming extra computational and transmission resources for the XR server. Similarly, within the requested FoV, background tiles can also be proactively dropped, since they can be rendered at the XR device using {\color{blue}diffusion}-based inpainting techniques  based on received surrounding tiles. However, for specific scenarios, given the differentiated computing capabilities of the devices, a trade-off between the computing latency and the transmission latency should be carefully studied. 
    \item \textbf{Foreground tile}: {\color{blue}Foreground tiles are used to represent interactive elements or objects that are closer to the player's viewpoint.} Concerning the foreground tiles with objects of interest to users, our focus lies in generating a significance map of the tiles based on the user requests. This map serves as a guide for semantic-aware delivery, as discussed in section~\ref{sec:IIIC}. Given that users exhibit varying levels of attention to the same object and different levels of attention to distinct objects, we first propose the derivation of a personalized significance map. This map relies on the matrix product of UOA values across all semantic features and the distribution of semantic features within each tile in the requested FoV~\cite{du2023attention}. Meanwhile, in a multicasting scenario, tiles requested by a greater number of users should also receive higher significance, given that the loss of them may affect multiple users at the same time. Thus, the ultimate significance map for all transmitted tiles is derived through the Hadamard product of the aggregated multi-user significance map corresponding to their requested FoV and the overlapping degree of the FoV requested by all users. The implications of the multi-user tile significance map on semantic-aware delivery are further explored in section~\ref{sec:IIIC}.
\end{itemize}

\subsection{Multi-user Data Delivery}
\label{sec:IIIC}
In this subsection, we propose to integrate semantic awareness into the online transcoding strategy as well as multicast delivery, to diminish pixel-level redundancy. 
\subsubsection{ Designing a semantic-aware multicast delivery scheme for enhancing experience and playback synchronization}
\begin{figure*}
    \centering \includegraphics[width=0.95\linewidth]{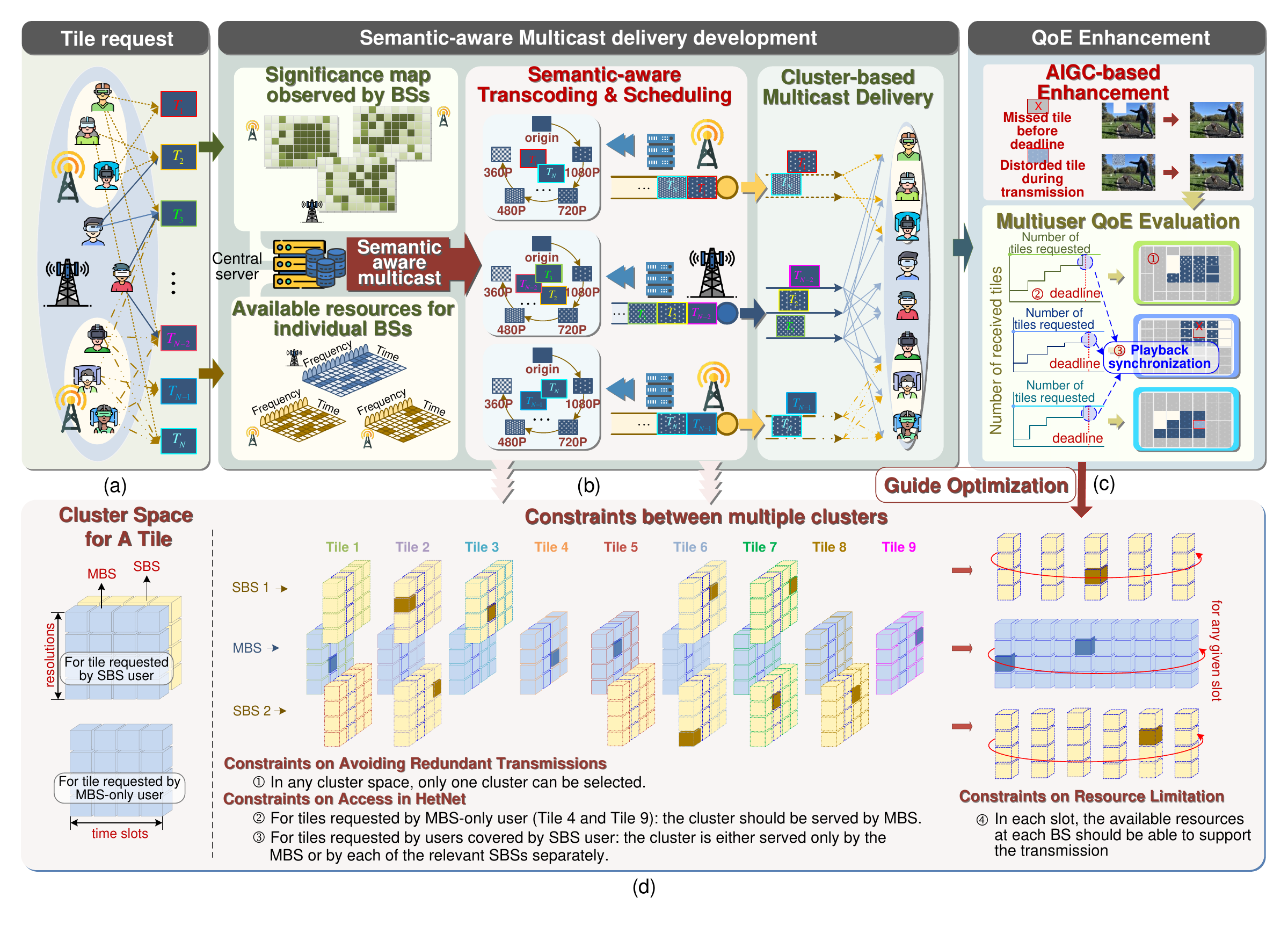}
    \caption{Tile significance map based
semantic-aware XR content
delivery with GAI collaborative
refinements. (a) System model for HetNet; (b) Illustration for semantic-aware multicast delivery; (c) GAI-based collaborative refinement and performance evaluation; (d) Optimization problem formulation for semantic-aware multicast delivery.}
    \label{fig:task3}
\end{figure*}

In this work, we consider a typical HetNet architecture with one macro BS (MBS) and several small BSs (SBSs) as shown in Fig.~\ref{fig:task3}(a). All the BSs assume a synchronized discrete-time system with slots. In HetNet, the SBSs generally share the same segment of frequency bandwidth and the MBS operates at a different frequency bandwidth. Meanwhile, all the SBSs can be covered by the MBS,  and there is no overlap between SBSs. In the considered multicast system, the resource allocation targets are no longer users but rather tiles requested in individual BSs. In this context, we define a multicast transmission about a tile as a cluster, which is specified by the associated BS, the transcoded resolution, and the transmission slot as shown in Fig.~\ref{fig:task3}(b). During the delivery, the following three types of constraints are considered.  
\begin{itemize}
    \item To avoid redundant transmission, a tile within a given BS can only be transmitted at most once. 
    \item For tiles requested by the user only covered by the MBS, the cluster should be served by MBS. For tiles requested by users covered by SBSs, the cluster is either served only by the MBS or by each of the relevant SBSs separately.
    \item In each slot, the available resources at each BS should be able to support the transmission.
\end{itemize}
For guiding the delivery optimization as illustrated in Fig.~\ref{fig:task3}(d), we propose a three-dimensional experience evaluation metric, as depicted in Fig.~\ref{fig:task3}(c). The first is the weighted resolution based on the {\color{blue}tile semantic significance acquired in Stage II}. The second is the playback smoothness, {\color{blue}which is jointly determined by transmission and computing latency}. The last is playback synchronization, which can minimize the time-shift between multiple users to ensure an immersive viewing experience for all the users in the virtual world. However, in achieving the optimal experience, the scheduling, transcoding, and multicast decisions are coupled with each other, which makes the optimization problem intractable. Therefore, we propose a possible way to decouple the high-dimensional optimization problem, which can unfold along three  steps: 
\begin{itemize}
    \item \textbf{Decoupling of Multicast}:  The reason for starting with this lies in two distinctive roles of MBS in HetNet. One is the only choice for users uncovered by SBSs, which restricts the clusters associated with these users to be served by MBS in priority. The other is the capability of simultaneous transmission to users covered by different SBSs, which suggests that the clusters related to the tiles requested in different SBSs should be preferably served by the MBS. In addition, it is desirable to balance the ratio of total significance scores to available bandwidth between BSs in order to keep the tiles with the same significance in different BSs to be transcoded into a similar resolution level.  
    \item \textbf{Decoupling of Transcoding}:  After Step 1, the coupling between BSs in the resource allocation is released. Then, the online transcoding  can performed within each BS separately based on their significance map. With the objective of the total weighted resolution, a natural idea is to give higher resolution level to the more important tile. However, specific decision-making scenarios still need to be determined taking into account the resources available at the BS.
    \item \textbf{Decoupling of Scheduling}: At last, scheduling decisions can be determined by jointly considering the significance map and the transmission progress of the tiles requested by individual users, with the aim of guaranteeing the playback synchronization between multiple users. Moreover, for scenarios with a high degree of dynamic change in wireless resources, the available resources for each time slot also need to be considered in the design of the algorithm.
\end{itemize}

 % At last, scheduling decisions can also be determined by jointly considering the significance map and the transmission degree of completion of tiles requested by each user, with the aim to guarantee the playback synchronization between multiple users. Moreover, for scenarios with a high degree of dynamic change in wireless resources, the available resources for each time slot also need to be considered in the design of the algorithm. In this case, the prediction of the available resources of the communication system helps to further complete the field scheduling optimization. Meanwhile, in this way, even if some tiles cannot be transmitted within the specified latency deadline, the discarded tiles are to be unimportant tiles, which can be remedied by GAI technology.
 \begin{figure*}
    \centering \includegraphics[width=1\linewidth]{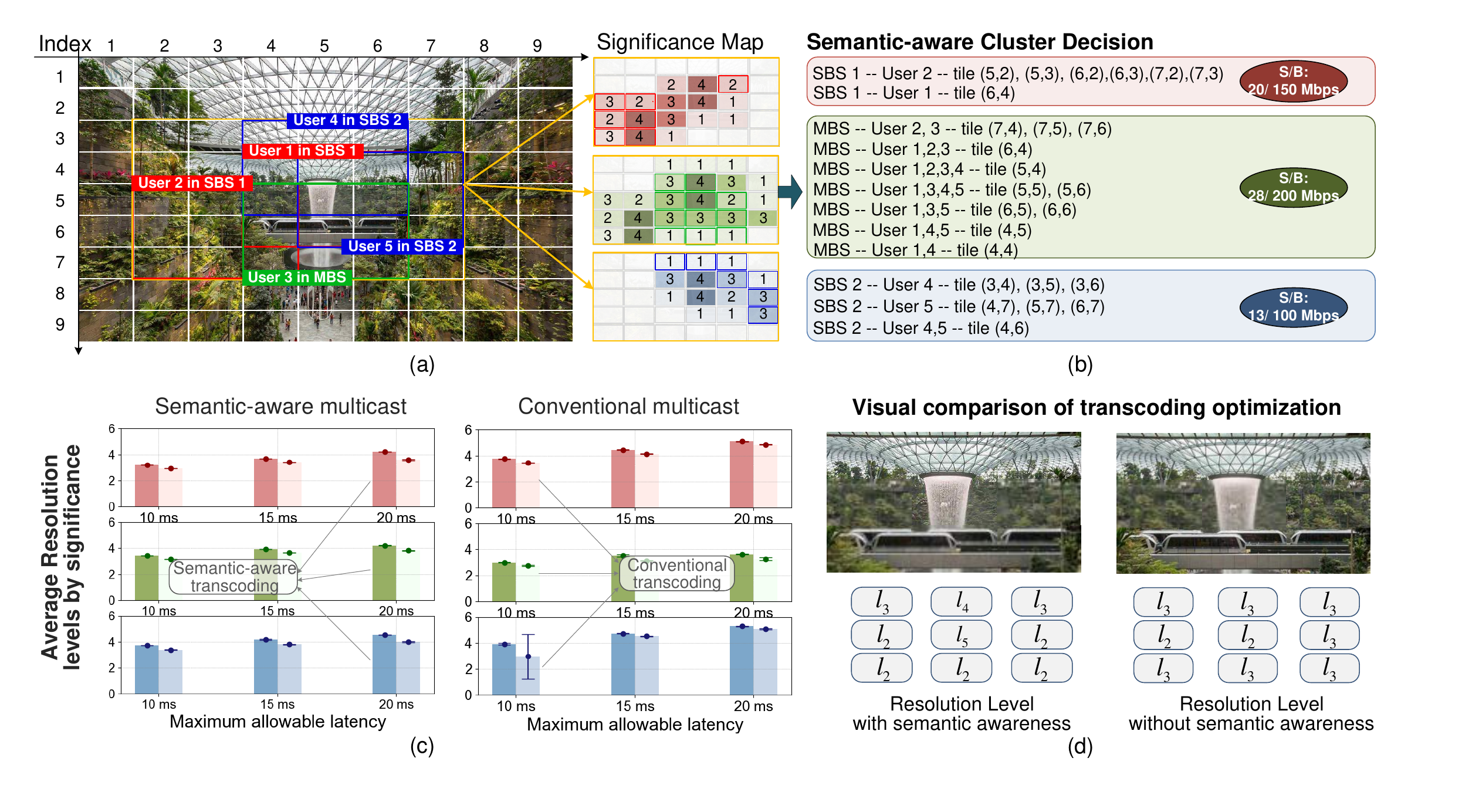}
    \caption{Implementation of semantic-aware multicast-based delivery strategy in case study. (a) Image of real scenes, Jewel Changi Airport and significance map; (b) Results of semantic-aware multicast decision; (c) Performance comparison under semantic-aware and conventional multicast with semantic-aware and conventional transcoding; (d) Visual comparison of semantic-
aware and conventional transcoding optimization, where  $l_5$ to $l_1$ represent the original resolution, 75\%, 50\%, 25\% and 15\% original resolution, respectively. }
    \label{fig:casestudy}
\end{figure*}
\subsubsection{Advancing experience with GAI-assisted denoising, FoV Completion and personalized enhancement}
During wireless transmission, transmitted data are susceptible to noise interference, resulting in distorted data that compromises the immersive experience. Furthermore, resource limitation may occasionally prevent the successful transmission of all the requested tiles to the user before playback, leading to playback lag or incomplete construction of virtual scenes.   Consequently, exploring corresponding remedies becomes imperative.

Given diffusion powerful generation capabilities and flexible deployment, it emerges as the primary choice for enhancing experience. The forward process of the diffusion model initiates with an input data sample and progressively introduces noise, mirroring the interference experienced by data transmitted in a wireless channel. Conversely, in the reserve process, starting with the noisiest data version, the model iteratively denoises the data, aiming to recover the original data, with a parallel objective to the denoising process in communication systems. While the literature showcases notable denoising capabilities for images{\color{blue}~\cite{xie2023diffusion}}, the determination of the minimal number of required denoising steps for specific noisy data  warrants further investigation to reduce computing latency. Simultaneously, thanks to the success of inpainting  techniques, the lost tile can be recovered.  {\color{blue}Considering the dual factors of computational latency and generation of tile fidelity, GAN and diffusion model can be chosen on a case-by-case basis to make a trade-off between the both factors.}
In addition to the compensation for transmission losses, GAI can open avenues for personalized user experiences, such as real-time voice translation and personalized ambiance rendering.

% \cite{jiang2023qoe, mu2023exploiting, zhu2022semantic}
% \cite{du2023attention, samant2022framework}
% \cite{yu2023inpaint,zhang2023ummaformer,lugmayr2022repaint,fan2023hierarchical,shilova2023adbooster, wang2023learning,du2023ai}
% \section{ Envision GeSa-XRF: Future Directions}
% \subsection{model training}

% \subsection{GAI computing reduction}
% \subsection{LLM and GAI}
\section{Case Study: Streamlined Multicast-based data delivery in GeSa-XRF }

In this case study, we focus on the implementation of semantic-aware multicast-based image delivery process in GeSa-XRF, specifically concerning semantic-aware multicast decision and semantic-aware transcoding, {\color{blue}which  determines the source content to be transmitted as well as the transmission bandwidth. The multi-modal SemCom can be implemented  following the results of the strategy.  }
\subsection{Scenario Setup}
Without loss of generality, we consider a HetNet with one MBS, two SBSs, and five users. Specifically, each SBS covers two users and one user is only covered by the MBS. For a given XR session, we develop the delivery strategy based on the image of real scenes, Jewel Changi Airport, shown in Fig.~\ref{fig:casestudy}(a). According to the overlap of the FoVs and the common semantic feature that people have across FoVs, we derive the semantic significance maps for each BS, respectively. The tiles highlighted in the three maps refer to the tiles requested by the covered users.  The darker the color of the tile, the more important it is. To ensure the playback smoothness, all the requested tiles should be delivered to the users within an allowable latency. In the subsequent development of the strategy, we assume the available bandwidth for the MBS, SBS 1, and SBS 2 to be 200~Mbps, 150~Mbps, and 100~Mbps, respectively. 
\subsection{Strategy Development and Performance Analysis}
Following the decoupling steps outlined in Section~\ref{sec:IIIC}, we begin by identifying the cluster associated tiles requested by user 3 (covered solely by the MBS), designating them to be served by the MBS. 
 To validate our proposed framework, we introduce two cluster grouping methods: semantic-aware multicast and conventional multicast.  In the conventional approach, our goal is to achieve load balance, that is, making the ratio of the number of tiles to be sent to each BS to the available bandwidth as equal as possible.  Conversely, in the semantic-aware approach, we strive for the similar ratio of the total importance score of the tiles to be sent and the available bandwidth across different BSs, to ensure that the tiles with the same significance to be transcoded into the similar resolution level. The semantic-aware multicast decision results can be found in Fig.~\ref{fig:casestudy}(b).
 Both the approaches can be implemented based on conditional linear programming with constraints. After the cluster decision, we perform the transcoding strategy for each BS, respectively. Similarly, we consider the two methods: semantic-aware and conventional transcoding. In semantic-aware transcoding, we aim to optimize the total resolution level weighted by the significance value, while in conventional transcoding, we just optimize the total resolution, regardless of the significance. To solve the above two combinatorial optimization problems, we adopt the genetic algorithm. Fig.~\ref{fig:casestudy}(c) indicates that average resolution levels by significance under semantic-aware transcoding are generally higher than those under conventional transcoding. 
{\color{blue} The metrics of PSNR and LPIPS achieved by the semantic-aware and semantic-unaware transcoding are 29.175 dB and 28.934 dB, 0.379 and 0.421, respectively.}
 Visual comparisons of the two transcoding optimization methods are presented in Fig.~\ref{fig:casestudy}(d). 
Furthermore, through a comparison of simulation results between semantic-aware multicast and conventional multicast, we observe a higher similarity in average resolution levels by significance across the three BSs in semantic-aware multicast. This observation establishes a basis for rationalizing multicast distribution of XR content in heterogeneous networks with multi-BS cooperation.
\section{Conclusion and Future Work}
In this paper, we have conducted a thorough exploration of efficient wireless XR transmission technologies. Specifically, we have seamlessly integrated semantic awareness to optimize data transmission, leveraging GAI for collaborative refinements and proposed a novel framework called GeSa-XRF. In the future work,  in the data collection stage, we intend to involve a more in-depth exploration about {\color{blue}the integration of the cutting-edge technologies in 6G, e.g., massive MIMO,} in the superposition scheme. During the data analysis stage, our focus will be on optimizing the number of proactively rendered and dropped tiles , taking into account the communication and computational resources {\color{blue}at cloud-edge-end architecture in 6G}. In data delivery stage, we will explore the optimal transmission strategy, concurrently considering three key experience evaluation metrics. Additionally, we aim to investigate more efficient decoupling strategies for high-dimensional optimization problems leveraging the {\color{blue}task reasoning capabilities of large language model techniques}.

\bibliographystyle{IEEEtran}
\bibliography{ref}
\end{document}